\documentstyle[12pt]{article}
\begin{document}

{\rightline August 15, 1996}

\begin{center} {\Huge Gauge theories induced by bosons in fundamental
representation}\\
\vskip0.6in A. Pap and P. Suranyi\\ Department of Physics, University of Cincinnati,
Cincinnati, Ohio 45221
\end{center}
\vfill
\begin{abstract}
 A lattice theory of scalar bosons in the fundamental representation of the gauge
group $SU(N_c)$  and of the global symmetry group $SU(N_f)$ is shown to induce a
standard gauge theory only at large
$N_f$.  The system is in a deconfined phase at strong scalar self-coupling and any
finite $N_f$.  The requirement of convergence of the effective gauge action imposes a
lower limit on the scalar mass.
\end{abstract} \vfill\eject 
\section{Introduction} There is ample numerical evidence, but still no analytic proof
for confinement in nonabelian gauge theories.   Lattice simulations of pure gauge
theories clearly show that large Wilson loops follow an area law,
$\langle W\rangle\sim e^{-cA}$, which is usually taken as the signal for
confinement.  Besides the scale parameter (or coupling constant) a nonabelian gauge
theory contains only one parameter, the number of colors,
$N_c$.   There were numerous attempts to explore simplifications which might occur
at the large
$N_c$ limit and allow for analytic solutions.  Confinement in lattice gauge theories
can be proven at strong coupling but in the relevant,  weak coupling, scaling, region
no analytic approach has succeeded. 

Kazakov and Migdal~\cite{migdal} conjectured that a gauge theory induced by a self
adjoint scalar field might be exactly solvable in the
$N_c\rightarrow\infty $ limit.  An induced gauge theory has a simpler structure
because its bare lagrangian does not contain the kinetic and self-interaction terms
of the gauge fields, familiar from QCD.  The hope was that after integrating out the
scalar bosons one ends up with an effective gauge theory, which is in the same
universality class as a standard lattice gauge theory.  The pure gauge terms of the
lagrangian are supposed to be generated by  scalars loops.  The theory would become
solvable due to the existence of a "master" scalar field.   It has been pointed out,
however  that the lagrangian of this model possesses an extra gauge symmetry,
which makes the expectation value of the standard Wilson loop vanish
identically~\cite{Gross}.  Such a property shows that unlike in QCD the 
quark-antiquark potential is infinite.  Quark-antiquark systems are pointlike. 
 
Various solutions have been offered to the problem of extra gauge symmetry. "Filled
Wilson loops" (combinations of loops in which every link variable and its conjugate
appear the same number of times) as the measure of confinement and the
perturbative (in variable $1/N_c$) introduction of pure gauge terms have been
proposed~
\cite{semenoff}. 

A way to eliminate the extra gauge symmetry is to introduce scalar fields in the
fundamental representation~\cite{ilchev}\cite{arafeva}.   The lattice lagrangian for a
scalar field with compact gauge interaction  is linear in the gauge field,
$U$, and does not have the extra gauge symmetry of the lagrangian for adjoint
representation fields.  Simulations have been performed with interacting fields in
mixed representations~\cite{ilchev} and the large $N_c$ limit has been
investigated~\cite{arafeva}.  
 
Our aim is to investigate an induced gauge theory with fundamental scalars in the
$N_f$ dimensional fundamental representation of a global flavor group $SU(N_f)$. 
Rather then taking the limit
$N_c\rightarrow\infty$, we will try to use analytic approximations at finite $N_c$. 
This approach is similar to that of Hasenfratz and Hasenfratz~\cite{hasenfratz}, who
investigated a gauge theory induced by fermions in the fundamental representation,
where the lagrangian is also linear in the compact gauge field.  One of their
conclusion, based on perturbative calculations, was that both in the continuum and
lattice theories
 a minimal number of fermions,
$N_f>11 N_c/2$, is needed to achieve asymptotic freedom in the induced gauge
theory.  They also pointed out that in the limit of large fermion mass one needs to
take the limit of $N_f\rightarrow\infty$ to be able to obtain a conventional lattice
gauge theory.  

Using fundamental scalars instead of fermions makes the problem more complex, due
to the self-interaction of scalars, which is required by renormalization theory.  As
we will see later the scalar self coupling has a significant effect on the phase of the
induced gauge theory. 

A perturbative one loop level analysis shows that the induced theory is
asymptotically free only if the number of flavors, $N_f$, satisfies the inequality
$N_f>22N_c$.  This condition is very similar to that of Hasenfratz and
Hasenfratz~\cite{hasenfratz} in fermion induced gauge theories~\cite{semenoff2}.
The main purpose of the present paper is to investigate confinement, or rather the
conditions under which the induced gauge theory is in the same universality class
with a standard gauge theory with a Wilson plaquette action.

\section{Lattice gauge theory with fundamental scalars}  In the absence of the pure
gauge (plaquette) term the lattice lagrangian density for $N_c$ scalars in the
fundamental representation of the gauge group,
$SU(N_c)$ and of the global flavor symmetry group, $SU(N_f$ has the following form:
\begin{equation}  L=m^2\phi^{\dagger}_i(x)\phi^i(x)+
\lambda_1\left[\phi^{\dagger}_i(x)\phi^i(x)\right]^2+\lambda_2
\left|\phi^{\dagger}_i(x)\phi^j(x)\right|^2+
\kappa^2\sum_{n=1}^{2D}\phi^\dagger_i(x+\hat e_n) U_n(x)\phi^i(x).
 \label{lagr} 
\end{equation} where the summation is over all  $2D$  unit lattice vectors (positive
for
$n\leq D$ and negative for $D< n\leq 2D$) originating at
$x$. Note that $U_n(x)=U^{\dagger}_{n+4}(x+\hat e_n)$, where $n=1,...,D$.  Summation
over gauge indices has been omitted.  $i$ and $j$ represent flavor indices.   We will
use a rescaled version of this Lagrangian, redefining the scalar fields $\phi(x)$ as
$m\phi(x)\rightarrow\phi(x)$.  We will also introduce rescaled couplings
$\lambda_i/m^2\rightarrow\lambda_i$.  This rescaling will amplify the role of the
large $m$ limit that will be repeatedly used below.   Note that one can only expect a
confined induced theory in the momentum range below the scalar mass.  $m$ serves
as a cutoff in the theory~ \cite{semenoff2}.  We will also see that the condition of
convergence of the effective gauge lagrangian imposes a lower bound on $m/\kappa$,
as well. 

 We will use a transformation to make the lagrangian quadratic in the scalar field. 
Introducing sources $j_1(x)$ and
$j_2(x)$ where the first is a gauge scalar, while the second is a hermitian gauge
matrix, we can write the generating functional as 
 \begin{eqnarray}  Z&=&\exp\left\{-\lambda_1\sum_x\frac{\delta^2} {\delta
j_1(x)^2} -\lambda_2{\rm{Tr}}\sum_x\frac{\delta^2} {\delta j_2(x)^2}\right\} \int
d\phi(x) dU_i(x)\nonumber\\ &&
\exp\Bigg\{ -\sum_x
\Big[\phi^\dagger(x)(j_1+j_2+1)\phi(x)+
\frac{\kappa^2}{m^2}\sum_{i=1}^{2D}\phi^\dagger(x+\hat e_i)
U_i(x)\phi(x)\Big]\Bigg\}.
 \label{gener}
\end{eqnarray}  Note that if $N_c>N_f$ then the tensor
$M_\beta^\alpha=\phi_i^\alpha\phi^{i\dagger}_\beta$, where $\alpha$ and
$\beta$ are gauge indices,  is degenerate, consequently ${j_2}^\alpha_\beta$ also
satisfies some constraints.  In what follows,  we will assume that $N_f\geq N_c$.

The integral over the scalar field, being Gaussian, can be performed to give
\begin{equation} Z=\exp\left\{-\lambda_1\sum_x\frac{\delta^2} {\delta j_1(x)^2}
-\lambda_2{\rm{Tr}}\sum_x\frac{\delta^2} {\delta j_2(x)^2}\right\}
\int dU(x) d^{-N},
\label{gener2}
\end{equation} where we dropped the subscript $f$ from $N_f$.

Now the determinant, $d$, can be rewritten using Schwinger's proper time formalism
as
\begin{equation} d={\rm Det}\left[j(x)+j_0(x)+1+
\frac{\kappa^2}{m^2}U(x)\right]^{-N}=\exp\left\{ N{\rm Tr}\int \frac{{\rm
d}T}{T}e^{-T[j(x)+j_0(x)+1+\frac{\kappa^2}{m^2}U(x)]}
\right\}.
\label{det}
\end{equation} Here $U$ is a step operator, connecting nearest neighbor lattice points.
$U$ satisfies the following constraint: $\langle x|U|x'\rangle=\langle
x'|U^\dagger|x\rangle$.  Note that the singularity at $T=0$ is an irrelevant constant,
independent of $U$ and
$j_i$.  After expanding the exponential in (\ref{det}) and transforming the operator
products into functional integrals~\cite{strassler} one obtains the following
representation for the generating  functional
 \begin{eqnarray} Z&=& \int dU(x)\sum_{n=0}^\infty\frac{N^n}{n!}
\exp\left\{-\lambda_1\sum_x\frac{\delta^2} {\delta j_1(x)^2}
-\lambda_2{\rm{Tr}}\sum_x\frac{\delta^2} {\delta j_2(x)^2}\right\}
\prod_{i=1}^n\int \frac{{\rm d}T_i}{T_i}e^{-T_i}\int {\rm d} x^{(i)}(\tau_i)
\nonumber\\ &&{\rm Tr}\exp\left\{-\int {\rm
d}\tau_i\left[j_1(x^{(i)}(\tau_i))+j_2(x^{(i)}(\tau_i)) +\frac{\kappa^2}{m^2}
U(x_i(\tau_i))\right]\right\},
 \label{newz}
\end{eqnarray} where $T_i$ is the length of the $i$th path.  Each of the path integrals
is path ordered in the $SU(N_c)$ space.  The $n$ paths are given in parametric
representation, $x^{(i)}(\tau_i)$, satisfying the condition $x^{(i)}(T_i)= x^{(i)}(0).$

Each term of the sum over $n$ in  (\ref{newz}) contains $n $ path integrals over
closed loops.  We will evaluate these functional integrals below.  Notice that while
the operators $j_i(x)$ are diagonal in coordinate space, operator  $U$ generates
discrete steps on the lattice.  For a finite size loop there can only be a finite number
of discrete steps. Only a finite power of the combination  $(\kappa^2/m^2)U$ will
contribute.  Expanding one of the functional integrals of (\ref{newz}) we obtain
\begin{eqnarray}  &&\int {\rm d} x(\tau){\rm Tr}\exp\left\{-\int {\rm
d}\tau\left[j_1(x(\tau))+j_2(x(\tau)) +\frac{\kappa^2}{m^2}
U(x(\tau))\right]\right\}=\sum_k\left(-{\rm
d}\tau\frac{\kappa^2}{m^2}\right)^k\nonumber\\&&
\sum_{{\cal L}_k}\sum_{\tau_i}{\rm
Tr}\left[\exp\left\{-\tau_0\left[j_1(x_0)+j_2(x_0)\right]\right\}\prod_{i=1}^k
U_{i-1i}\exp\left\{-\tau_i\left[j_1(x_i)+j_2(x_i)\right]\right\}\right],
\label{functional}
\end{eqnarray} where summation over ${\cal L}_k$ denotes all possible lattice paths
of length
$k$ lattice units.  Summation over $\tau_i$, the length of the
$T$-interval between two subsequent operators, $U_{i-1i}$ and $U_{ii+1}$, can be
turned into integrals over $\tau_i$,  $\int\prod_{i=1}^k{\rm d}\tau_i$. The variables
$\tau_i$ satisfy the relation
$\sum_{i=0}^k\tau_i=T$.  Introducing a multiplier $1=\int{\rm
d}\tau_0\delta(\tau_0+\tau_1+...+\tau_k-T)$ we can see that the integrand depends
only on
$\tau_0+\tau_k$. Integrating over $\tau_0$ introduces a multiplier
$\tau_0+\tau_k$.  After the substitution
$\tau_0+\tau_k
\rightarrow\tau_k$ and adding all cyclic permutations of the operators $U_i$ we
obtain for (\ref{functional})
\begin{equation}  T\sum_k\left(\frac{\kappa^2}{m^2}\right)^k
\sum_{{\cal L}_k}\int \prod_{i=1}^n{\rm d}{\tau_i}\delta\left(T-\sum
\tau_i\right){\rm Tr}\left[\prod_{i=1}^k
U_{i-1i}\exp\left\{-\tau_i\left[j_1(x_i)+j_2(x_i)\right]\right\}\right].
\label{functional2}
\end{equation} The multiplier -1 of
$\kappa^2/m^2$ was omitted because $k$, the number of links of a closed loop  is
always even. 

Notice now that $\exp\{-\tau [j_1+j_2]\}$ are translation operators for the
functional derivatives with respect to the sources $j_1$ and $j_2$.  After
translating the differentiations and setting $j_i=0$  in (\ref{newz})  we obtain extra
terms in the effective action.  These extra terms represent contact path-path
interactions.  The
$U$ matrices are path ordered and traced along $n$ different paths.  The contact
interaction proportional to $\lambda_1$ is scalar, but the other one, proportional to
$\lambda_2$ is proportional to operator $P_{ij}$ that crosses the $SU(N_c)$ indices
of  path $i$ and path $j$ at the point of intersection.  The integration over
$T$ can be performed using the delta-functions in (\ref{functional2}) to result in the
following expression for the generating function
\begin{eqnarray} Z&=& \int
dU_i(x)\sum_{n=0}^\infty\frac{N^n}{n!}\sum_{r_1}\sum_{r_2}...
\sum_{r_n}\left(
\frac{\kappa^2}{m^2}\right)^{\sum_k{r_k}}\sum_{{\cal L}_{r_1}}\sum_{{\cal
L}_{r_2}}...\sum_{{\cal L}_{r_n}}
\nonumber\\&&\prod_{k=1}^n\left[ {\rm Tr}\prod_{j=1}^{r_k}\int{\rm d}\tau^{(k)}_j
\exp\left\{-\tau^{(k)}_j\right\}
 U_{jj+1}^{(k)}\right]\prod_{k,l=1}^n\exp\left\{-\sum_{i,j}
\delta_{x_i^{(k)},x_j^{(l)}}\tau_i^{(k)}\tau_j^{(l)}
[\lambda_1+P_{ij}\lambda_2]\right\}.
 \label{newestz}
\end{eqnarray} The $j$th vertex of the $k$th loop is denoted by $x_j^{(k)}$. 
$\tau_j^{(k)}$ denotes the "time" (fraction of $T$)   spent at that vertex.  The
partition function is a sum over the products of
$n$ closed loops of perimeter $r_k$. ${\cal L}_{r_k}$ symbol is used to identify
specific loops of perimeter $r_k$.  In other words, ${\cal L}_{r_k}$ defines the shape
and location of the loop. The last multiplier in (\ref{newestz}) is a loop-loop
interaction term generated by the scalar self-interaction. 

 If the scalar couplings vanish then the last multiplier of (\ref{newestz}) is 1 and the
integrations over
$\tau_i^{(k)}$ and the summation over $n$ can be performed to result in  the
exponentiated form
\begin{equation} Z= \int dU_i(x)\exp\left\{N\sum_{r}\left(
\frac{\kappa^2}{m^2}\right)^r\sum_{{\cal L}_{r}}{\rm Tr}\prod_{j=1}^{r}
U_{jj+1}\right\}.
\label{noint}
\end{equation} This is exactly the form of the partition function one would obtain in a
much simpler way, namely by expanding the logarithm of the boson determinant in a
power series of 
$(\kappa^2/m^2)U$ and summing over cyclic permutations of the product
${\rm Tr}\prod_{j=1}^{r} U_j$.

A necessary condition for the convergence of the expansion can easily be obtained. 
Take an arbitrary closed loop.  Now split the loop open at any of its vertices, $x_0$,
and add an "appendix" consisting of a set of $k$
$U$-matrices on links, $(x_i,x_{i+1})$, $i=0,...,k-1$, ending up at vertex $x_k$.  Close
the loop by adding all the $k$ links $(x_i,x_{i-1})$ to end up at $x_0$.  In other
words, the new loop, obtained after this surgery,  is doubled up on the segment
between $x_0$ and $x_k$. Due to pairwise cancellations the corresponding
contribution of
$U$ matrices is
$(\kappa^2/m^2)^{2k}U_{x_0,x_1}...U_{x_{k-1,k}}U_{x_k,x_{k-1}}...U_{x_1,x_0}=
(\kappa^2/m^2)^{2k}$.  Consequently one obtains a contribution of the same form as
that of the unsplit loop.  Clearly, the expansion of the action in (\ref{noint})
converges only if the sum over all the possible additions of the above kind converges. 
Since the doubled up path is a random walk of length
$k$, the number of these walks is $(2D)^k$ the condition for convergence is
$m^2>\sqrt{2D}\kappa^2$.  This condition gets modified if $\lambda_i\not=0$.  Then
no exact condition can be obtained, because self-intersections of the random walks
between
$x_0$ and $x_k$ alter the weight.   Neglecting the effect of self-intersections of the
doubled up line the condition for convergence get modified as
\begin{equation} m^2>\sqrt{2D}\kappa^2 I_2(\lambda_1+N_c\lambda_2),
\label{condit}
\end{equation} where
\begin{equation} I_2(x)=\int_0^\infty{\rm d}\tau_1{\rm
d}\tau_2e^{-\tau_1-\tau_2-\tau_1
\tau_2 x}=\frac{1}{x}e^{\frac{1}{x}}Ei\left(\frac{1}{x}\right).
\label{intersec}
\end{equation} The origin of the notation $I_2$ and further properties of this
function will be discussed later.  Note that every projection operator $P$ generates a
multiplier $N_c$ by creating one more closed gauge index loop along the doubled up
path. The function
$I_2(x)$ satisfies the condition $I_2(x)\leq1$.  $I_2(0)=1$. The coefficients of the
weak coupling expansion are all finite but the expansion, having a zero radius of
convergence, is only asymptotic. 
$I_2(x)\simeq (1/x)\log x$ if
$x$ is large.  The physical reason for the decreasing nature of $F(x)$ is clear: the
 scalar self-interaction and with that the path-path interaction  is also repulsive,
making the excursion from
$x_0$ to
$x_k$ energetically less and less favorable.  In the limit of
$\lambda_i\rightarrow\infty$ doubled up loops are forbidden.  In general, the larger
$\lambda$ is, the better the convergence of the expansion.

 A few more comments about (\ref{newestz}) are in order. The contribution
$k=l$ and $i=j$ renormalizes $m$ and can be omitted.  This can be seen in the
representation of the partition function without rescaling the scalar field,
$m\phi\rightarrow\phi$.  The contributions
$i\not=j$ or/and $k\not=l$ of the contact term prevent us from resumming the series
over $n$.   The summation can only be performed using a linked cluster expansion.  
One expects the induced gauge theory to be QCD-like only at momentum scales below
$m$, when the only role of the scalars to generate a gauge coupling,  because free
bosons cannot be produced~\cite{semenoff2}.

Let us now examine the effect of path-path interactions more closely.  Our aim is to
generate an effective action using a linked cluster decomposition. We will call a
certain set of $n$ lattice loops a diagram.  Two loops are connected if they have at
least one common vertex.  A subdiagram is connected if there is a continuous path
formed from links of its loops between every pair of its vertices.  The contribution
of connected diagrams to the effective action can be calculated  in the following
manner.  Suppose that there are $s_m$ vertices in the diagram where $m$ loops
intersect.  A point where
$m$ loops intersect (including self intersections) contributes to the diagram by
\begin{equation} I_m=\int \prod_{i=1}^m \left[{\rm d}\tau_i e^{-\tau_i} \right]
\exp\left\{-\sum_{i\not=j}^m(\lambda_0+P_{ij}
\lambda_1)\tau_i\tau_j\right\}
\label{Im}
\end{equation}

Then the contribution of a connected diagram consisting of $n$ loops of perimeter
$r_k$, $k=1,...,n$ to the effective action  equals
\begin{equation} S=N^n\left(\frac{\kappa^2}{m^2}\right)^{\sum_kr_k}
\prod_{k=1}^n {\rm Tr}\prod_{j=1}^{r_k}
 U_{jj+1}^{(k)}W^{(1,...,k)},
\label{contr}
\end{equation} where 
\begin{equation} W^{(1,...,k)}=\prod_m (I_m)^{s_m}
\label{weight}
\end{equation} is the weight factor appearing due to self and mutual intersections of
$k$ connected loops. 

 The linked cluster expansion has the following form
\begin{equation} S=\sum_{k=1}^\infty\frac{N^k}{k!}\sum_{r_1,r_2,...,r_k}
\left(\frac{\kappa^2}{m^2}\right)^{r_1+r_2+...+r_k}\sum_{{\rm Conn}[{\cal L}_{r_1}
{\cal L}_{r_2}...{\cal L}_{r_k}]}\prod_{i=1}^k{\rm Tr}\left[\prod_{j=1}^{r_1}
U_{jj+1}^{(i)}\right]X^{(1,2,...,k)},
\label{cluster}
\end{equation} where Conn[${\cal L}_{r_1} {\cal L}_{r_2}...{\cal L}_{r_k}]$ means that
summation is over connected diagrams only.  The coefficients
$X^{(1,2,...,k)}$ satisfy the following recursion relation
\begin{equation} W^{(1,...,k)}=k!\sum_{m_1,...,m_k}\delta(k-m_1-2m_2-...-km_k)
\prod_{i=1}^k
\frac{1}{m_i!}\left[\frac{X^{(1,2,,...,i)}}{i!}\right]^{m_i}
\label{recursion}
\end{equation} We can solve these recursion relations iteratively to arrive at the 
following expression for
$k=1,2,$ and
$3$:
 \begin{equation} X^{(1)}=W^{(1)},
\label{cluster1}
\end {equation}
\begin{equation} X^{(1,2)}=W^{(1,2)}-W^{(1)}W^{(2)},
\label{cluster2}
\end {equation}
\begin{equation}
X^{(1,2,3)}=W^{(1,2,3)}+2W^{(1)}W^{(2)}W^{(3)}-W^{(1)}W^{(2,3)}-W^{(2)}W^{(1,3)}
-W^{(3)}W^{(1,3)}.
\label{cluster3}
\end {equation} Of course, the last three terms of $X^{(1,2,3)}$ lead to identical
terms in $S$.

At infinite coupling one only has self avoiding walks.  Then the overlap functions
$W^{(1,...,k)}=0$ for
$k>1$ and $W^{(1)}=1$.  The recursion relation (\ref{recursion}) can be solved exactly
to give the following weights
\begin{equation} X^{(1,...,k)}=\frac{(-1)^{k-1}}{k}.
\label{solved}
\end{equation} 

The worthiness of the cluster expansion depends on the value of the scalar couplings.
In fact, for large couplings the cluster coefficients, $X^{(r)}$, tend to a finite
constant. Contributions for diagrams having multiple overlaps decrease very rapidly. 
The overlap function
$W^{(1,...,m)}$ has the following asymptotic behavior:
\begin{equation}
 W^{(1,...,m)}\sim\frac{ \left(\log\lambda\right)^{s_2}}{\lambda^t},
\label{wbeh}
\end{equation} where $t=\sum_{m=2}ms_m/2$. Thus, it is advantageous to use
expansion with respect to $W^{(1,...,r)}$.  If the coupling is small, however, then the
cluster expansion is more appropriate, because
$W^{(1,...,r)}=1-O(\lambda),$ while the asymptotic behavior of cluster coefficients is
\begin{equation} X^{(1,...,r)}\sim\lambda^{r-1}.
\label{xbeh}
\end{equation}

\section{ Strong scalar self coupling}

At infinitely strong scalar self coupling the magnitude  of the scalar field is fixed. It
was shown ~\cite{Frohlich} that a pure scalar self interacting theory on a lattice at
infinite coupling is equivalent to a self avoiding random walk.  In our system, as
shown by (\ref{newestz}), the Wilson loops in the partition function avoid each other
and themselves (note that trivial self interaction of lattice points has been
previously removed).  If no intersection is allowed the integration over $\tau_i^{(k)}$
can be trivially performed. One obtains for the
 partition function
\begin{equation} Z= \int
dU_i(x)\sum_{n=0}^\infty\frac{N^n}{n!}\sum_{r_1}\sum_{r_2}...
\sum_{r_n}\left(
\frac{\kappa^2}{m^2}\right)^{\sum_k{r_k}}\sum_{{\cal L}_{r_1}}\,\!'\sum_{{\cal
L}_{r_2}} \,\!'...\sum_{{\cal L}_{r_n}}\,\!'\prod_{k=1}^n{\rm
Tr}\prod_{i=1}^{r_k}U_i^{(k)},
\label{infinite}
\end{equation} where the primes over the summation signs imply summation over
non-intersecting loops only.  If intersections are not allowed then each gauge
variable $U_i^{(k)}$ appears only once under the integral and the integral vanishes for
all $n$ except for $n=0$. Then the
 partition function itself is 1.  The expectation value of  a Wilson loop,
$\prod_1^S U_i$ is 
\begin{equation}
\langle\prod_1^S U_i\rangle=N\left(
\frac{\kappa^2}{m^2}\right)^S.
\label{loop1}
\end{equation} This is clearly a perimeter law, implying that there is no confinement
in the
$\lambda\rightarrow\infty$ limit.  The system is in a deconfined phase.

 The system remains in a deconfined phase at a large enough value of the scalar
coupling constants.  If the coupling is strong we can estimate the expectation value
of the loop by employing an expansion in
$1/\lambda$, using the asymptotic behavior of the intersection weights, 
$W(\lambda)$.  They behave roughly as $1/\lambda^{r/2}$ for a $r$-fold intersection,
as shown by (\ref{wbeh}).  Then the leading contribution to the expectation value of a
Wilson loop of area $A$ and perimeter $S$ is still (\ref{loop1}).  Contributions of
more then 1 loop, having $r$ intersections are cut off by the $r$th power of
$I_2\kappa^2/m^2\simeq(\log\lambda/\lambda)(\kappa^2/m^2).$ In particular, the
contribution coming from single plaquette loops only is
\begin{equation}
\langle{\rm Tr} \prod_{i=1}^SU_i\rangle\simeq
...+\left[N\left(I_2(\lambda)\right)^2\left(\frac{\kappa^2}{m^2}\right)^4\right]^A,
\label{cont3}
\end{equation} clearly negligible compared to the main contribution, (\ref{wbeh}), if
$N$ is kept finite at large values of
$\lambda$.    Also, corrections to this large $\lambda$ expansion converge rapidly. 
The conclusion is that the system is in a deconfined state at sufficiently strong
scalar coupling. If $N$ is allowed to increase as well then the situation may change,
as discussed in the following section. 

\section{Large $N$ limit}

Let us consider first $\lambda_i=0$, for simplicity. If $\lambda_i\rightarrow0$, than
the coefficients
$I_m=1$ and the integrand of the partition function exponentiates.  Loops of all sizes
contribute as shown by (\ref{noint}).  It is easy to estimate the contribution of loops
of different sizes in the effective action to the expectation value of the Wilson loop
if $\kappa^2/m^2$ is small.  A loop coinciding with the Wilson loop contributes by the
same as (\ref{loop1}).  $A$ plaquettes, where $A$ is the area of the loop contribute by
\begin{equation}
\langle\prod_1^S U_i\rangle=N^A\left(
\frac{\kappa^2}{m^2}\right)^{4A},
\label{loop2}
\end{equation} a negligible contribution, unless $N$ is also large.  If $N$ is small
then at vanishing self coupling of the bosons the Wilson loop follows
 a perimeter law.  The system is in a deconfined state.   

As shown by (\ref{noint}) we recover the standard Wilson action if we take the
simultaneous limit of
$m^2\rightarrow\infty$, $N\rightarrow\infty$, such that
$1/g^2=N(kappa^2/m^2)^4$ tends to a finite value~\cite{arafeva}\cite{hasenfratz}.  If
$m^2
$ is not infinite, just large, then one has the area law only for sufficiently small
loops:  
\begin{equation}
\frac{A}{S}< \frac{\log(m^2/\kappa^2)}{\log g^2}.
\label{condition}
\end{equation} where $A$ is the area of the loop.  We will return to the case of finite
$N$ and
$\kappa^2/m^2$ in the next section.

In the rest of this section we will investigate a theory obtained from (\ref{newestz})
by taking the limits $N\rightarrow\infty$,
$m^2\rightarrow\infty$ at fixed 
$g^2=m^2/N\kappa^2$.   We will allow for arbitrary $\lambda_i $.  Then only
contributions containing  plaquettes survive.  The contribution of a connected
diagram, consisting of two plaquettes, to the effective action is
\begin{equation} S_2=\left[(I_2)^{s_2}-1\right] \left(\frac{1}{g^2}\right)^2{\rm
Tr}U_{P_1} {\rm Tr}U_{P_2},
\label{twoplaq}
\end{equation} where $s_2$ is the number of points where the two plaquettes
intersect (1,2, or 4).  The multiplier $(I_2)^{s_2}-1$ vanishes at $\lambda_i=0$ and
goes to -1 if $\lambda_i\rightarrow\infty$. 

Let us examine the effect of small couplings, $\lambda_i$.  The cluster functions
$X^{(1,...,r)}$  decrease as $\lambda^{r-1}$. Then obviously two plaquette clusters
dominate.  One can substitute a pair of plaquettes with a cluster of two plaquettes in
the covering of Wilson loop approximately $A$ different ways.  There are four
different relevant 2 plaquette clusters, two of which have one common point,
$s_2=1$, and two have two common points,
$s_2=2$.  Altogether one has obtains a correction term $A[2I_2+2I_2^2-4]$.
Exponentiating the correction term leads to a similar correction term to the area law
at vanishing $\lambda$.  At small $\lambda$ the correction term is just $-6\lambda
A$.  Since the correction term is negative, the confining linear  potential gets
stronger  at increasing values of
$\lambda$. Clusters of larger size will not change this conclusion, because the
cluster functions are proportional to higher powers of
$\lambda$ and their sum also exponentiates to an area law.  

In principle, the cluster expansion could diverge at some finite value of $\lambda$
and fixed $g$.  Then the system could go into a deconfined phase.  It is easy to see,
however, that this is not the case.  Examine the theory
 at large $\lambda$. Then again it is not convenient to use the cumulant expansion. 
Though the gauge coupling might be small, in effect one can use a strong coupling
expansion, because the overlap of plaquettes is costly, each one introducing an
multiplier of $I_2\sim \log\lambda/\lambda$.  Then the $A$ plaquette contribution
dominates, just as at large gauge coupling, giving the following estimate for the
expectation value of a Wilson loop:
\begin{equation}
\langle {\rm Tr}\prod_{i=1}^S U_i\rangle\sim
\left(\frac{I_4(\lambda)}{g^2}\right)^A.
\label{largel}
\end{equation} For a minimal covering an $I_4$ overlap functions appears at every
internal lattice point of the Wilson loop.  This is so because there are four adjacent
plaquettes at every point. Since $I_4\sim 1/\lambda^2$,  (\ref{largel}) provides a
good approximation if $g^2\lambda^2>>1$.  The conclusion is that the system is in the
confined phase at any fixed  $g$ provided $\lambda>>1/g$.   Since we saw earlier that
it is in a confined phase at small $\lambda$ as well, it follows that the limit
$m^2\rightarrow\infty$, $g^2$ fixed leads to a confining theory at all values of
$\lambda$.  When $\lambda\rightarrow\infty$ the string constant tends to infinity
and we approach a superconfined theory, similar to the one induced by adjoint
representation bosons.

\section{Conclusions}

We have investigated the phases of an $SU(N_c)$ lattice gauge theory induced by
$N$ complex bosons.  At the perturbative level such a theory is asymptotically free if
the the number of flavors, $N$, satisfies $N>22N_c$~\cite{hasenfratz}.  The
convergence of the induced theory was shown to  require a minimum value for the
scalar mass.  Also, on physical grounds it is expected that a sensible induced gauge
theory is obtain at energy scales below the scalar mass, which plays the role of a
cutoff.  For these reasons the scalar mass was assumed to be large.  

 The self-coupling, $\lambda$, of scalars, inducing the theory plays an important
role.  For large $\lambda$  and finite but large scalar mass (i.e., fixed "magnitude" for
the scalars) the system was shown to be in a deconfined state.   An obvious limit
where a confining theory might be obtained is $N, m^2\rightarrow\infty$,
$N(\kappa^2/m^2)^4$ fixed.  We show that both at small and large
$\lambda$ one obtains a confining theory (provided the standard lattice gauge theory
is confining).  Then it follows that the limits 
$m^2\rightarrow\infty$ and $\lambda\rightarrow\infty$ do not commute.  

It is of considerable interest to determine the phase of the system at finite but large
$m$, finite $\lambda$ and $N$.  We have a rather roundabout argument that, at least
at some values of the parameters, the induced gauge theory has a confined phase. The
effective gauge coupling (the coefficient of the plaquette action) is
$1/g^2\sim N\kappa^2/m^2$, but there are terms in the effective action corresponding
loops larger then plaquettes. Therefore, it is not at all obvious whether the theory
confines.   However, if $m$ is sufficiently large, then at fixed $g$ coefficients of 
terms corresponding to larger loops decrease rapidly.  Imagine now that instead of an
induced gauge theory we consider a real gauge system with gauge coupling $g$.  The
only difference between the effective actions of the standard and induced theories is
a change in the coefficient of the plaquette term, $N(\kappa^2/m^2)^4\rightarrow
N(\kappa^2/m^2)^4+1/g^2$.  If parameters are such that this change is sufficiently
small, then  no change is expected in the phase of the system  because the ratio of
the plaquette term and terms corresponding to larger loops hardly changes.  A real
scalar-gauge boson system on a lattice is expected to have both confined and
deconfined phases and at sufficiently large bare scalar mass it will be in the
confined phase.  Ultimately, the phase of the system at finite values of the
parameters can only be determined by simulations. 

  \section*{ACKNOWLEDGEMENTS}

This paper has been supported in part by the U.S. Department of Energy under grant
number DE-FG02-84ER40153. The authors also thank  Dr. G. Semenoff for discussions.


\begin{thebibliography}{10}
\small
\addtolength{\itemsep}{-6pt}
\bibitem{migdal} V. Kazakov and A. Migdal, Nucl. Phys. B397 (1993) 214.
\bibitem{Gross} I. Kogan, G. Semenoff, and N. Weiss, Phys. Rev. Lett. 69 (1992) 3435;
D. Gross, Phys. Lett. 293B (1992) 181.
\bibitem{semenoff}  G.W. Semenoff, in "Trieste 1993, Proceedings, High energy
physics and cosmology" 379-401.
\bibitem{ilchev} A. Ilchev, M. Moodley, and A. Welter, Phys. Lett. B332 (1994) 387
\bibitem{arafeva} I. Ya. Arefeva, Phys.Lett B308 (1993) 347.
\bibitem{hasenfratz} A. Hasenfratz,  Nucl.Phys.Proc.Suppl.30 (1993) 801;  A.
Hasenfratz and P. Hasenfratz, Phys.Lett. B297 (1992) 166.
\bibitem{semenoff2} The authors thank G. Semenoff for a discussion on this subject. 
\bibitem{strassler} M. Strassler, Nucl.Phys. B385 (1992)145. 
\bibitem{Frohlich}  C. Arago de Carvalho, S. Caracciolo, and J. Froelich, Nucl. Phys.
B215[FS7] (1983) 209.

 \end{thebibliography}
\end{document}